\documentstyle[fleqn,espcrc2,epsf]{article}

\begin{document}
\flushbottom
\thispagestyle{empty}

\title{\vspace*{-.75in}
\font\fortssbx=cmssbx10 scaled \magstep1
\hbox to \hsize{
\includegraphics{uwlogo.ps}
\hskip.25in \raise.05in\hbox{\fortssbx University of Wisconsin - Madison}
\hfill$\vtop{\normalsize\hbox{\bf MADPH-97-1026}
                \hbox{November 1997}
                \hbox{\hfil}}$ }
Large Natural Cherenkov Detectors: Water and Ice\footnotemark}

\author{Francis Halzen\address{Physics Department, University of Wisconsin, Madison, WI 53706, USA}}

\begin{abstract}\noindent
In this review we first address 2 questions:
\begin{itemize}
\item 
why do we need kilometer-scale muon and neutrino detectors?
\item
what do we learn from the operating Baikal and AMANDA detectors about the construction of kilometer-scale detectors?
\end{itemize}
 I will subsequently discuss the challenges for building the next-generation detectors. The main message is that these are different, in fact less ominous, than for commissioning the present, relatively small, detectors which must reconstruct events far outside their instrumented volume in order to achieve large effective telescope area. 
\end{abstract}

\maketitle
\renewcommand{\thefootnote}{\fnsymbol{footnote}}
\footnotetext{Talk presented at the {\it 5th International Workshop on Topics in Astroparticle and Underground Physics (TAUP\,97)}, Gran Sasso, Italy, Sept.~1997.}

\section{Why Kilometer-Scale Detectors?}

High energy neutrino telescopes are multi-purpose instruments; their science mission covers astronomy and astrophysics, cosmology,  particle physics and cosmic ray physics. Their deployment creates new opportunities for glaciology and oceanography, possibly geology\cite{pr}. The observations of astronomers span 19 decades in energy or wavelength, from radio-waves to the high energy gamma rays detected with satellite-borne detectors\cite{turner}. Major discoveries have been historically associated with the introduction of techniques for exploring new wavelengths. The important discoveries were surprises. In this spirit, the primary motivation for commissioning neutrino telescopes is to cover uncharted territory: wavelengths smaller than $10^{-14}$~cm, or energies in excess of 10~GeV. This exploration has already been launched by truly pioneering observations using air Cherenkov telescopes\cite{weekes}. Larger cosmic ray arrays with sensitivity above $10^7$~TeV, an energy where charged protons may point back at their sources with minimal deflection by the galactic magnetic field, will be pursuing similar goals\cite{cronin}. Could the high energy skies be devoid of particles? No, cosmic rays with energies exceeding $10^8$~TeV have been recorded\cite{cronin}. Between GeV gamma rays and the most energetic cosmic rays, there is uncharted territory spanning some eight orders of magnitude in wavelength. Exploring this energy region with neutrinos does have the definite advantage that they can, unlike high energy photons and nuclei, reach us, essentially without attenuation in flux, from the largest red-shifts.

The challenge is that neutrinos are difficult to detect: the small interaction cross sections that enable them to travel without attenuation over a Hubble radius, are also the reason why kilometer-scale detectors are required in order to capture them in sufficient numbers to do astronomy\cite{halzenkm}. There is nothing magical about this result ---\break
 I will explain this next.

Cosmic neutrinos, just like accelerator neutrinos, are made in beam dumps. A beam of accelerated protons is dumped into a target where they produce pions in collisions with nuclei. Neutral pions decay into gamma rays and charged pions into muons and neutrinos. All this is standard particle physics and, in the end, roughly equal numbers of secondary gamma rays and neutrinos emerge from the dump. In man-made beam dumps the photons are absorbed in the dense target; this may not be the case in an astrophysical system where the target material can be more tenuous. Also, the target material may be light rather than nuclei. For instance, with an ambient photon density a million times larger than the sun, approximately $10^{14}$ per cm$^3$, particles accelerated in the superluminal jets associated with active galactic nuclei (AGN), may meet more \mbox{photons} than nuclei when losing energy. Examples of cosmic beam dumps are tabulated in Table~1. They fall into two categories. Neutrinos produced by the cosmic ray beam are, of course, guaranteed and calculable. We know the properties of the beam and the various targets: the atmosphere, the hydrogen in the galactic plane and the CMBR background. Neutrinos from AGN and GRBs (gamma ray bursts) are not guaranteed, though both represent good candidate sites for the acceleration of the highest energy cosmic rays. That they are also the sources of the highest energy photons reinforces this association.

\begin{table}
\caption{Cosmic Beam Dumps}
\bigskip
\centering
\tabcolsep=1em
\begin{tabular}{|c|c|}
\hline
\bf Beam& \bf Target\\
\hline
cosmic rays& atmosphere\\
cosmic rays& galactic disk\\
cosmic rays& CMBR\\
AGN jets& ambient light, UV\\
shocked protons& GRB photons\\
\hline
\end{tabular}
\end{table}

In astrophysical beam dumps, like AGN and GRBs, there is typically one neutrino and photon produced per accelerated proton\cite{pr}. The accelerated protons and photons are, however, most likely to suffer attenuation in the source before they can escape. So, a hierarchy of particle fluxes emerges with $\rm protons<photons<neutrinos$. A generic neutrino flux can be obtained from this relation by, conservatively, equating the neutrino with the observed cosmic ray flux. The detector size can now be determined\cite{halzenzas} by taking into account the detection efficiency for neutrinos which is much reduced compared to protons. 

Neutrino telescopes are conventional particle detectors which use natural and clear water and ice as the Cherenkov medium. A three dimensional grid of photomultiplier tubes maps the Cherenkov cone radiated by a muon of neutrino origin. Nanosecond timing  provides degree resolution of the muon track which is, at high energy, aligned with the neutrino direction. The probability to detect a TeV neutrino is roughly $10^{-6}$\cite{pr}. It is easily computed from the requirement that, in order to be detected, the neutrino has to interact within a distance of the detector which is shorter than the range of the muon it produces. In other words, in order for the neutrino to be detected, the produced muon has to reach the detector.
Therefore,
\begin{equation}
P_{\nu\to\mu} \simeq {R_\mu\over \lambda_{\rm int}} \simeq A E_{\nu}^n \,,
\end{equation}
where $R_{\mu}$ is the muon range and $\lambda_{\rm int}$ the neutrino interaction length. For energies below 1~TeV, where both the range and cross section depend linearly on energy, $n=2$. Between TeV and PeV energies $n=0.8$ and $A=10^{-6}$, with $E$ in TeV units. For EeV energies $n=0.47$, $A =10^{-2}$ with $E$ in EeV.

At PeV energy the cosmic ray flux is of order 1 per m$^{2}$ per year and the probability to detect a neutrino of this energy is of order 10$^{-3}$. A neutrino flux equal to the cosmic ray flux will therefore yield only a few events per day in a kilometer squared detector. At EeV energy the situation is worse. With a rate of 1 per km~$^2$ per year and a detection probability of 0.1, one can still detect several events per year in a kilometer squared detector provided the neutrino flux exceeds the proton flux by 2 orders of magnitude or more. For the neutrino flux generated by cosmic rays interacting with CMBR photons and such sources as AGN and topological defects\cite{schramm}, this is indeed the case. All above estimates are conservative and the rates should be higher because the neutrinos escape the source with a flatter energy spectrum than the protons\cite{halzenzas}. In summary, the cosmic ray flux and the neutrino detection efficiency define the size of a neutrino telescope. Needless to say that a telescope with kilometer squared effective area represents a neutrino detector of kilometer cubed volume.

\section{Baikal and the Mediterranean}

First generation neutrino detectors are designed to reach a relatively large telescope area and detection volume for a neutrino threshold of tens of GeV, not higher and, possibly, lower. This relatively low threshold permits calibration of the novel instrument on the known flux of atmospheric neutrinos.  Its architecture is optimized for reconstructing the Cherenkov light front radiated by an up-going, neutrino-induced muon rather than for detecting signals of TeV energy and above. Up-going muons are to be identified in a background of down-going, cosmic ray muons which are more than $10^6$ times more frequent for a depth of 1~kilometer.

The ``landscape" of neutrino astronomy is sketched in Table~2. With the termination of the pioneering DUMAND experiment, the efforts in water are, at present, spearheaded by the Baikal experiment\cite{domogatsky}. Operating with 144 optical modules (OM) since April 1997, the {\it NT-200} detector will be completed by April 1998. The Baikal detector is well understood and the first atmospheric neutrinos have been identified; we will discuss this in more detail further on. The Baikal site is competitive with deep oceans although the smaller absorption length requires somewhat denser spacing of the OMs. This does however result in a lower threshold which is a definite advantage, for instance in WIMP searches. They have shown that their shallow depth of 1 kilometer does not represent a serious drawback. By far the most significant advantage is the site with a seasonal ice cover which allows reliable and inexpensive deployment and repair of detector elements.

\begin{table}[t]
\caption{}
\epsfxsize=7cm\epsffile{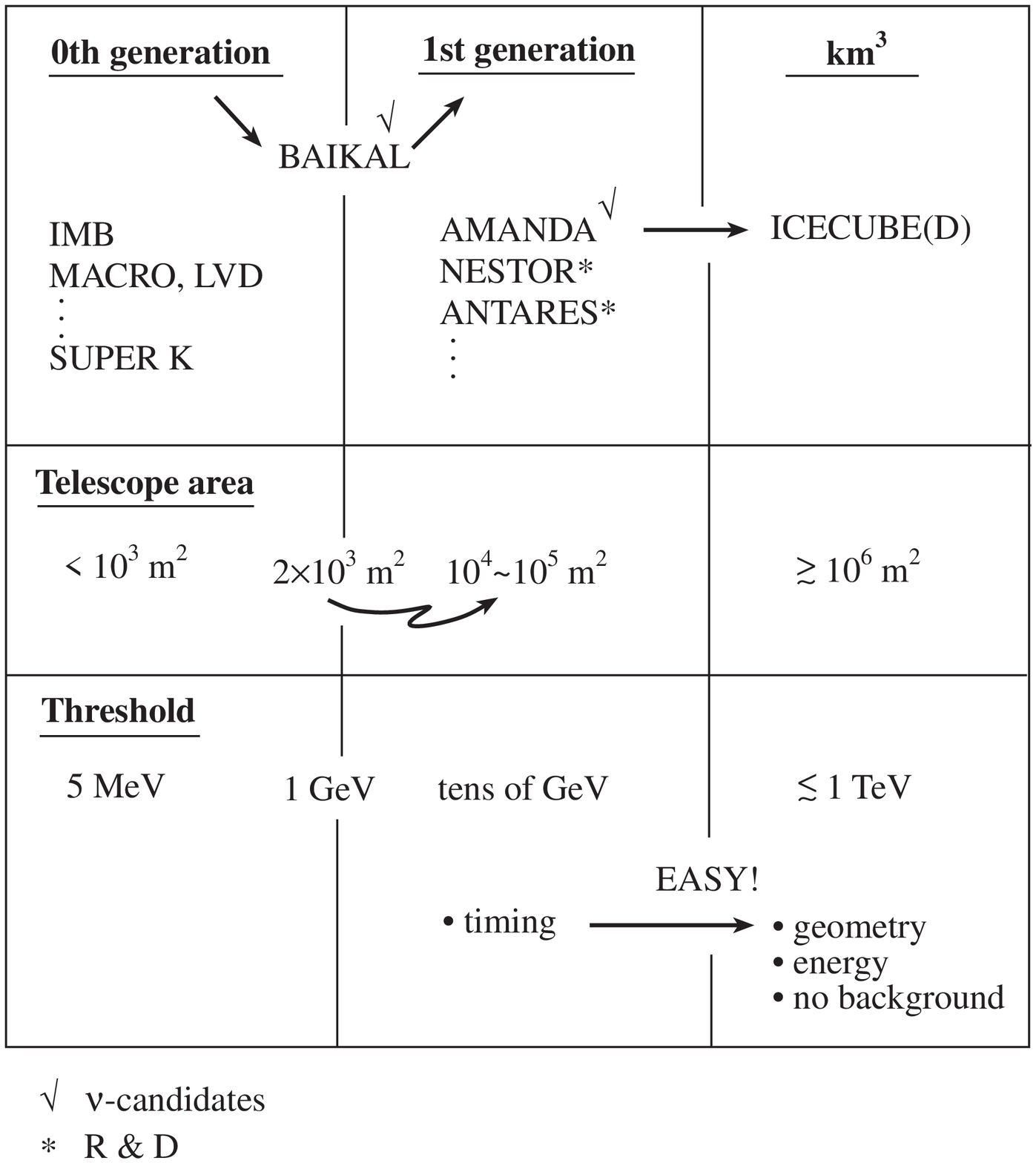}
\end{table}

In the following years, {\it NT-200} will be operated as a neutrino telescope with an effective area between $10^3 \sim 5\times 10^3$~m$^2$, depending on the energy. Presumably too small to detect neutrinos from AGN and other extraterrestrial sources, {\it NT-200} will serve as the prototype for a larger telescope. For instance, with 2000 OMs, a threshold of  $10 \sim 20$~GeV and an effective area of $5\times10^4 \sim 10^5$~m$^2$, an expanded Baikal telescope would fill the gap between present underground detectors and planned high threshold detectors of cube kilometer size. Its key advantage would be low threshold.

The Baikal experiment represents a proof of concept for deep ocean projects. These should have the advantage of larger depth and optically superior water. Their challenge is to design a reliable and affordable technology. Three groups are confronting the problem: NESTOR and Antares in the Mediterranean and a group at LBL, Berkeley. The latter has made seminal contributions to the design of digital OMs. Most of them have recently joined the AMANDA experiment and, as far as deployments are concerned, their effort will proceed in ice.

The NESTOR collaboration\cite{resvanis}, as part of an ongoing series of technology tests, has recently deployed two aluminum ``floors", 34~m in diameter, to a depth of 2600~m. Mechanical robustness was demonstrated by towing the structure, submerged below 2000~m, from shore to the site and back. The detector will consist of 12 six-legged floors separated by 30~m.

The Antares collaboration\cite{feinstein} is in the process of determining the critical detector parameters at a 2000~m deep, Mediterranean site off Toulon, France. A deliberate development effort will lead to the construction of a demonstration project consisting of 3 strings with a total of 200 OMs.

For neutrino astronomy to become a viable science several of these, or other, projects will have to succeed. Astronomy, whether in the optical or in any other wave-band, thrives on a diversity of complementary instruments, not on ``a single best instrument". When the Soviet government tried out the latter method by creating a national large mirror project, it virtually annihilated the field.

\section{First Neutrinos from Baikal}

The Baikal Neutrino Telescope is being deployed in Lake Baikal, Siberia, 3.6~km from shore at a depth of 1.1~km. An umbrella-like frame holds 8 strings, each instrumented with 24 pairs of 37-cm diameter {\it QUASAR} photomultiplier tubes (PMT). Two PMTs in a pair are switched in coincidence in order to suppress background from bioluminescence and PMT noise.

They have analysed 212 days of data taken in 94-95 with 36 OMs. Upward-going muon candidates were selected from about $10^{8}$ events in which more than 3 pairs of PMTs triggered. After quality cuts and $\chi^2$ fitting of the tracks, a sample of 17 up-going events remained. These are not generated by neutrinos passing the earth below the detector, but by showers from down-going muons originating below the array. In a small detector such events are expected. In 2 events however the light does not decrease from bottom to top, as expected from invisible showering muons below the detector. A detailed analysis \cite{ourneu} yields a fake probability of 2\% for both events.

After the deployment of 96 OMs in the spring of 96, three neutrino candidates have been found in a sample collected over 18 days. This is in agreement with the expected number of approximately 2.3.  One of the events is displayed in Fig.~1. In this analysis the most effective quality cuts are the traditional $\chi^2$ cut and a cut on the probability of non-reporting channels not to be hit, and reporting channels to be hit ($P_{\rm no\, hit}$ and $P_{\rm hit}$, respectively). To guarantee a minimum lever arm for track fitting, they were forced to reject events with a projection of the most distant channels on the track smaller than 35 meters. This does, of course, result in a loss of threshold.

\begin{figure}[t]
\centering\leavevmode
\epsfxsize=4cm\epsffile{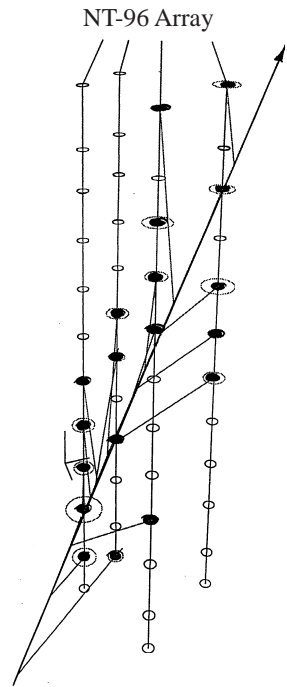}

\caption{Candidate neutrino event from NT-96 in Lake Baikal.}
\end{figure}

\section{ The AMANDA South Pole Neutrino Detector}

Construction of the first-generation AMANDA detector\cite{barwick} was completed in the austral summer 96--97. It consists of 300 optical modules deployed at a depth of 1500--2000~m; see Fig.~2. An optical module consists of an 8~inch photomultiplier tube and nothing else. Calibration of this detector is in progress, although data has been taken with 80 OM's which were deployed one year earlier in order to verify the optical properties of the ice (AMANDA-80).

\begin{figure}[t]
\centering
\hspace{0in}\epsfxsize=7.5cm\epsffile{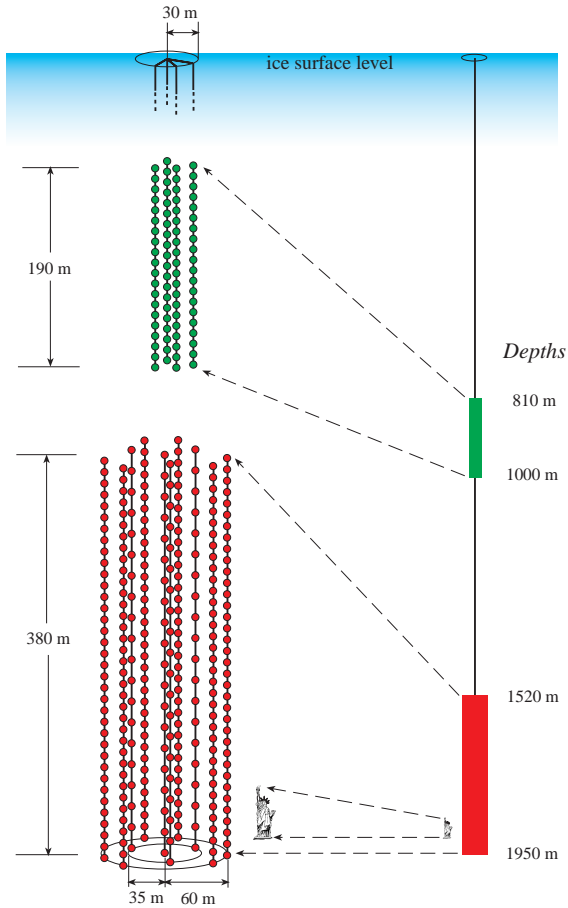}

\caption{The Antarctic Muon And Neutrino\break
 Detector Array (AMANDA).}
\end{figure}

\looseness=-1
The performance of the AMANDA detector is encapsulated in the event shown in Fig.~3. Coincident events between AMANDA-80 and four shallow strings with 80 OM's (see Fig.~2), have been triggered for one year at a rate of 0.1~Hz. Every 10 seconds a cosmic ray muon is tracked over 1.2 kilometer. The contrast in detector response between the strings near 1 and 2~km depths is dramatic: while the Cherenkov photons diffuse on remnant bubbles in the shallow ice, a straight track with velocity $c$ is registered in the deeper ice. The optical quality of the deep ice can be assessed by viewing the OM signals from a single muon triggering 2 strings separated by 79.5~m; see Fig.~3b. The separation of the photons along the Cherenkov cone is well over 100~m, yet, despite some evidence of scattering, the speed-of-light propagation of the track can be readily identified.

The optical properties of the ice are quantified by studying the propagation in the ice of pulses of laser light of nanosecond duration. The arrival times of the photons after 20~m and 40~m are shown in Fig.~4 for the shallow and deep ice\cite{serap}. The distributions have been normalized to equal areas; in reality, the probability that a photon travels 70~m in the deep ice is ${\sim}10^7$ times larger. There is no diffusion resulting in loss of information on the geometry of the Cherenkov cone in the deep ice.

\section{Intermezzo: AMANDA before\hfil\break and after}

The AMANDA detector was antecedently proposed on the premise that inferior properties of ice as a particle detector with respect to water could be compensated by additional optical modules. The technique was supposed to be a factor $5 {\sim} 10$ more cost-effective and, therefore, competitive. The design was based on then current information:

\begin{itemize}
\addtolength{\itemsep}{-2mm}
\item
the absorption length at 370~nm, the wavelength where photomultipliers are maximally efficient, had been measured to be 8~m,

\item
the scattering length was unknown,

\item
the AMANDA strategy was to use a large number of closely spaced OM's to overcome the short absorption length. Muon tracks triggering 6 or more OM's are reconstructed with degree accuracy. Taking data with a simple majority trigger of 6 OM's or more at 100~Hz yields an average effective area of $10^4$~m$^2$.
\end{itemize}

\noindent
The reality is that:
\begin{itemize}
\addtolength{\itemsep}{-2mm}
\item
the absorption length is 100~m or more, depending on depth\cite{science},

\item
the scattering length is $25 {\sim} 30$~m (preliminary),

\item
because of the large absorption length OM spacings are similar, actually larger, than that those of proposed water detectors. Also, a typical event triggers 20 OM's, not~6. Of these more than 5 photons are not scattered. In the end, reconstruction is therefore as before, although additional information can be extracted from scattered photons by minimizing a likelihood function which matches measured and expected \mbox{delays\cite{christopher}}.
\end{itemize}

\renewcommand{\thefigure}{\arabic{figure}a}
\begin{figure}[t]
\centering
\hspace{0in}\epsfysize=9cm\epsffile{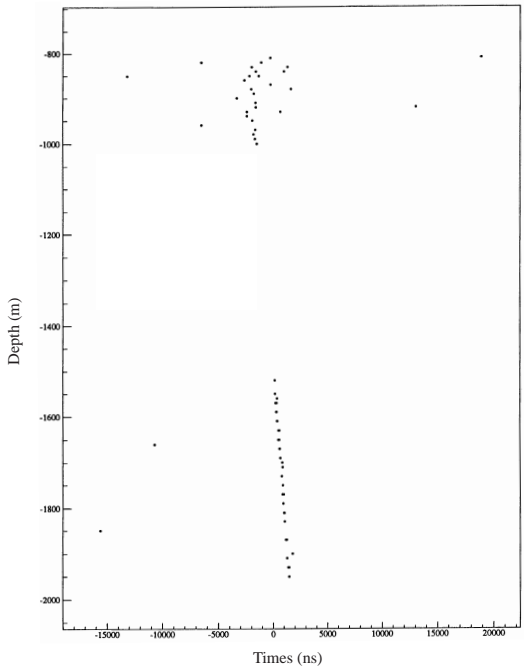}

\caption{Cosmic ray muon track triggered by both shallow and deep AMANDA OM's. Trigger times of the optical modules are shown as a function of depth. The diagram shows the diffusion of the track by bubbles above 1~km depth. Early and late hits, not associated with the track, are photomultiplier noise.}
\end{figure}

\addtocounter{figure}{-1}\renewcommand{\thefigure}{\arabic{figure}b}
\begin{figure}[t]
\centering
\hspace{0in}\epsfysize=9cm\epsffile{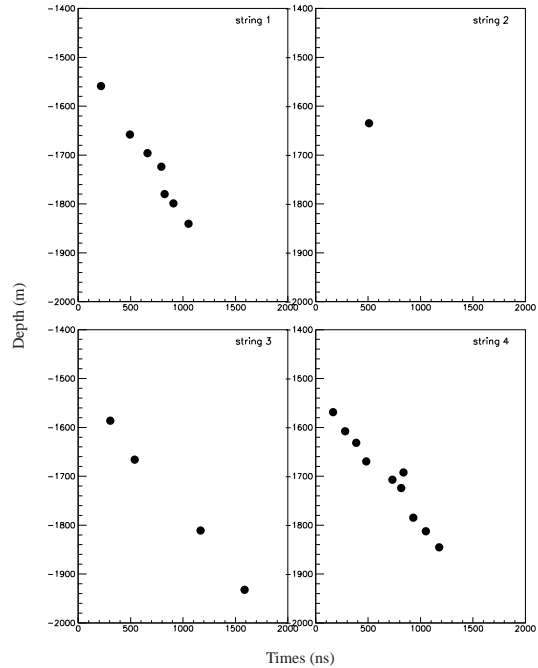}

\caption{Cosmic ray muon track triggered by both shallow and deep AMANDA OM's. Trigger times are shown separately for each string in the deep detector. In this event the muon mostly triggers OM's on strings 1 and 4 which are separated by 79.5~m. }
\end{figure}
\renewcommand{\thefigure}{\arabic{figure}}

\begin{figure*}[t]
\centering
\hspace{0in}\epsfxsize=14.5cm\epsffile{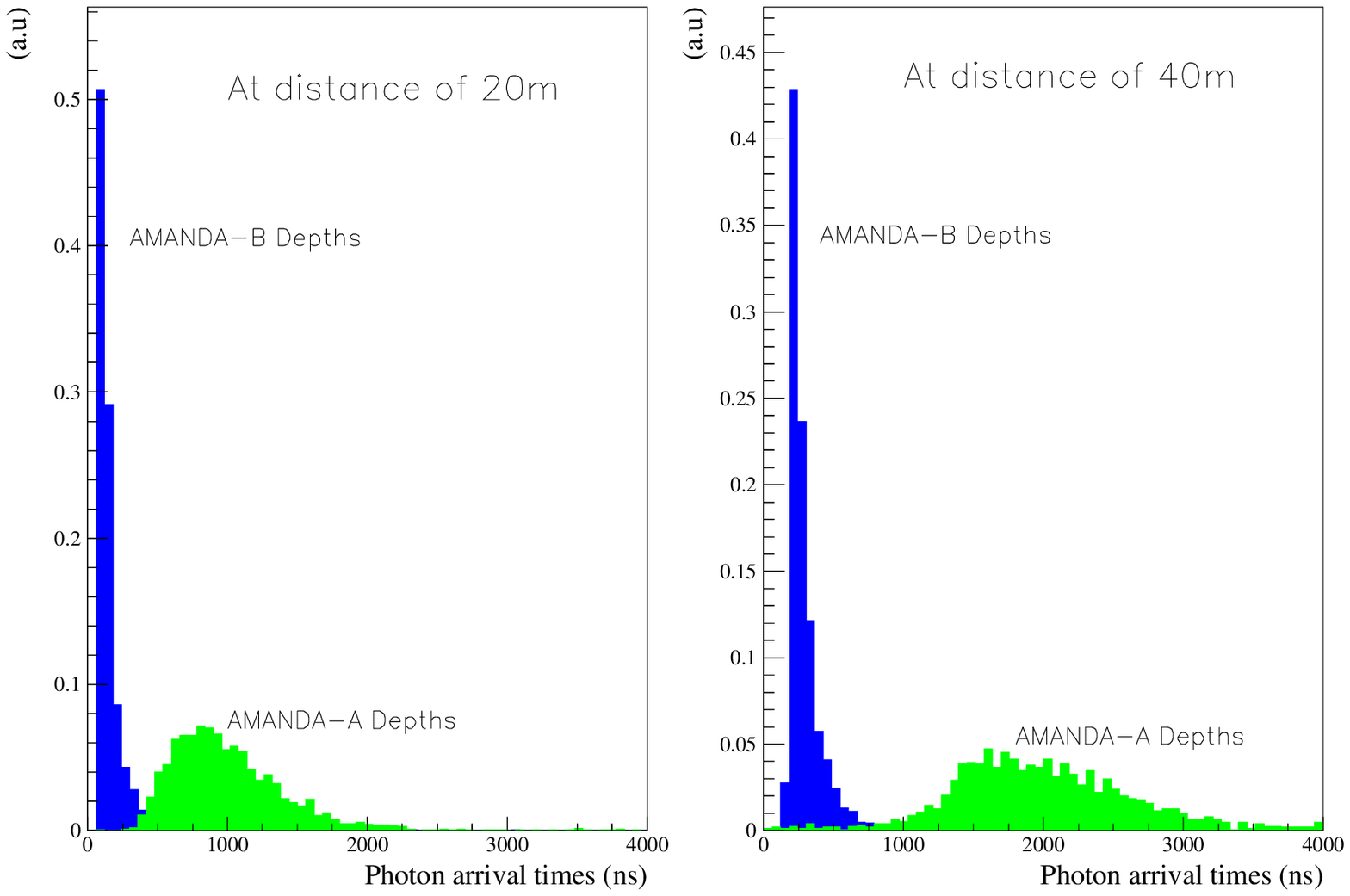}

\caption{Propagation of 510~nm photons indicate bubble-free ice below 1500~m, in contrast with ice with some remnant bubbles above 1~km.}
\end{figure*}

The measured arrival directions of background cosmic ray muon tracks, reconstructed with 5 or more unscattered photons, are confronted with their known angular distribution in Fig.~5. The agreement with Monte Carlo simulation is adequate. Less than one in $10^5$ tracks is misreconstructed as originating below the detector\cite{serap}. Visual inspection reveals that the remaining misreconstructed tracks are mostly showers, radiated by muons or initiated by electron neutrinos, which are reconstructed as up-going tracks of muon neutrino origin. At the $10^{-6}$ level of the background, candidate events can be identified; see Fig.~6. This exercise establishes that AMANDA-80 can be operated as a neutrino detector; misreconstructed showers can be readily eliminated on the basis of the additional information on the amplitude of OM signals. Monte Carlo simulation, based on this exercise confirms, that AMANDA-300 is a $10^4$~m$^2$ detector (somewhat smaller for atmospheric neutrinos and significantly larger for high energy signals) with 2.5 degrees mean angular resolution\cite{christopher}. We have verified the angular resolution of AMANDA-80 by reconstructing muon tracks registered in coincidence with a surface air shower array\cite{miller}.

\begin{figure*}
\centering
\leavevmode
\epsfxsize=14cm\epsffile{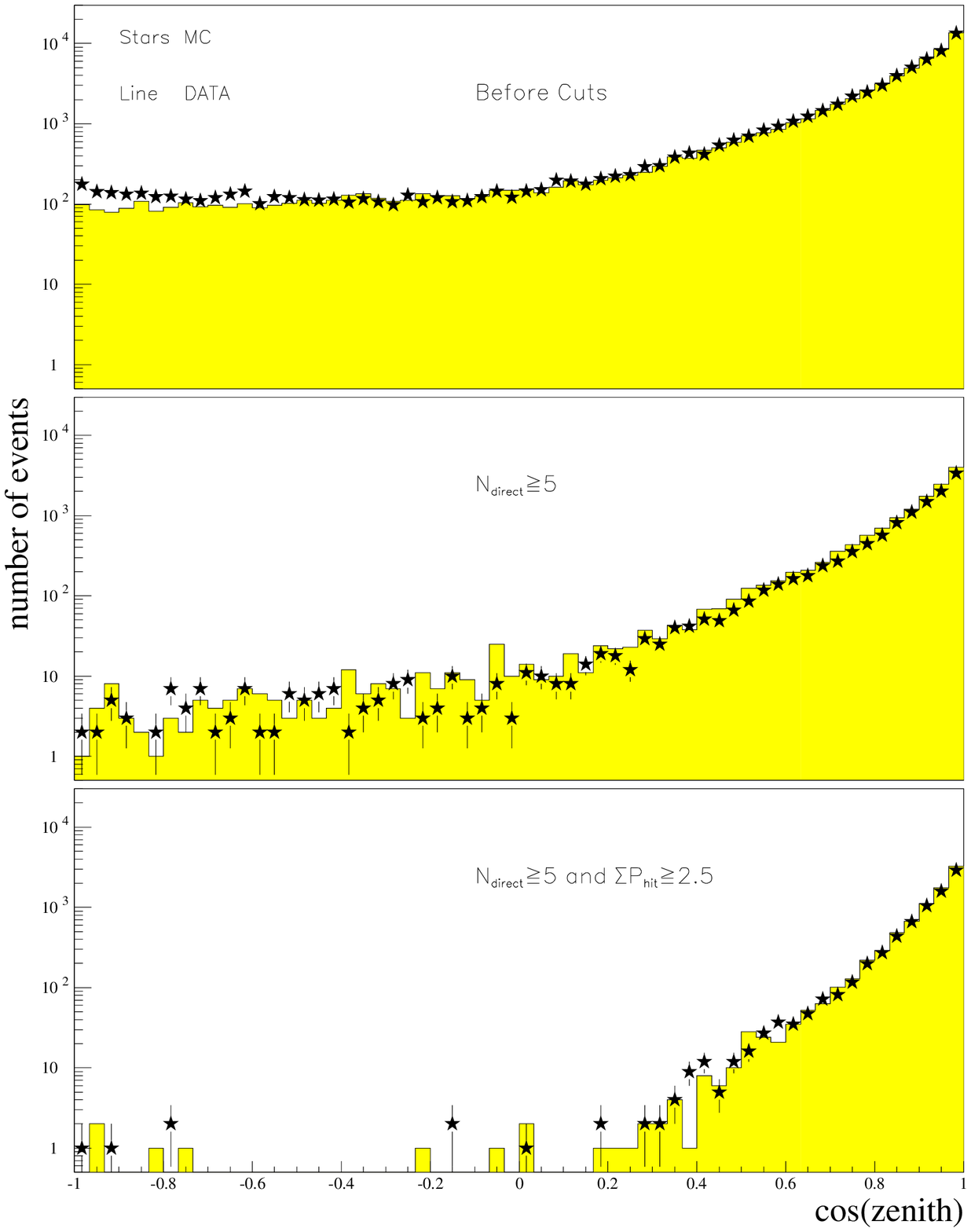}

\caption{Reconstructed zenith angle distribution of muons triggering AMANDA-80: data and Monte Carlo. The relative normalization has not been adjusted at any level. The plot demonstrates a rejection of cosmic ray muons at a level of 10$^{-5}$ which is reached by only 2 cuts: a cut on the number of unscattered photons and on $P_{\rm hit}$, a quantity introduced in the context of the Baikal experiment.}
\end{figure*}

\begin{figure}[t]
\centering
\hspace{0in}\epsfxsize=2.9cm\epsffile{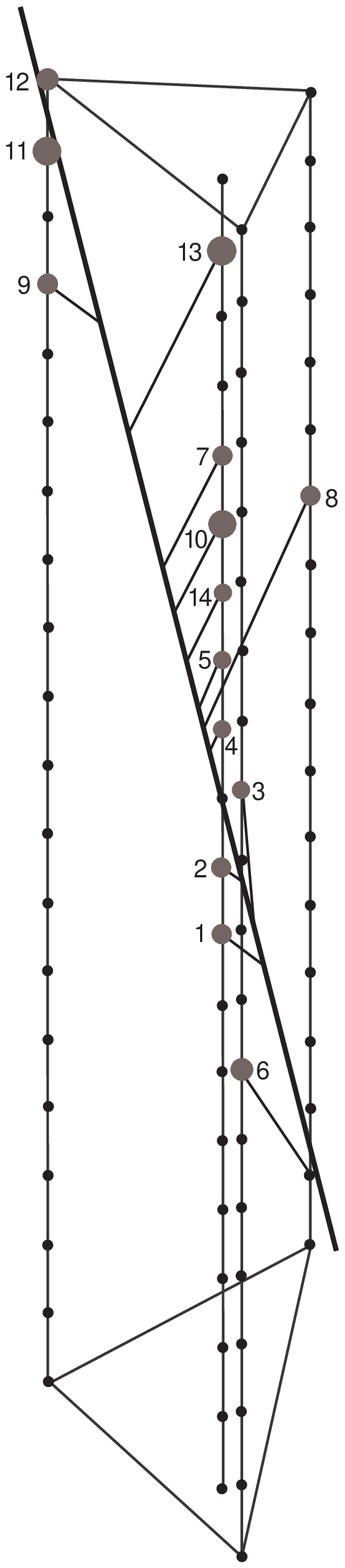}

\caption{A candidate up-going, neutrino-induced muon in the AMANDA-80 data. The numbers indicate the time sequence of triggered OMs, the size of the dots the relative amplitude of the signal.}
\end{figure}

\section{Towards ICE CUBE(D)}

A strawman detector with effective area in excess of 1~km$^2$ consists of 4800~OM's: 80 strings spaced by $\sim$~100~m, each instrumented with 60~OM's spaced by 15~m. A cube with a side of 0.8~km is thus instrumented and a through-going muon can be visualized by doubling the length of the lower track in Fig.~3a. It is straightforward to convince oneself that a muon of TeV energy and above, which generates single photoelectron signals within 50~m of the muon track, can be reconstructed by geometry only. The spatial positions of the triggered OM's allow a geometric track reconstruction with a precision in zenith angle of:
\begin{eqnarray}
&&\mbox{angular resolution} \simeq {\mbox{OM spacing}\over\mbox{length of the track}}\nonumber\\
 &&\hskip.5in\simeq \mbox{15\,m/800\,m} \simeq \mbox{1\,degree};
\end{eqnarray}
no timing information is really required. Timing is still necessary to establish whether a track is up- or down-going, not a challenge given that the transit time of the muon exceeds 2 microseconds.  Using the events shown in Fig.~3, we have, in fact, already demonstrated that we can reject background cosmic ray muons. Once ICE CUBE(D) has been built, it can be used as a veto for AMANDA and its threshold lowered to GeV~energy.

With half the number of OM's and half the price tag of the Superkamiokande and SNO solar neutrino detectors, the plan to commission such a detector over 5 years is not unrealistic. The price tag of the default technology used in AMANDA-300 is \$6000 per OM, including cables and DAQ electronics. This signal can be transmitted to the surface by fiber optic cable without loss of information. Given the scientific range and promise of such an instrument, a kilometer-scale neutrino detector must be one of the best motivated scientific endeavors ever.

\section{About Water and Ice}

The optical requirements of the detector medium can be readily evaluated, at least to first order, by noting that string spacings determine the cost of the detector. The attenuation length is the relevant quantity because it determines how far the light travels, irrespective of whether the photons are lost by scattering or absorption. Remember that, even in the absence of timing, hit geometry yields degree zenith angle resolution. Near the peak efficiency of the OM's the attenuation length is 25--30~m, larger in deep ice than in water below 4~km. The advantage of ice is that, unlike for water, its transparency is not degraded for blue Cherenkov light of lower wavelength, a property we hope to take further advantage of by using wavelength-shifter in future deployments.

The AMANDA approach to neutrino astronomy was initially motivated by the low noise of sterile ice and the cost-effective detector technology. These advantages remain, even though we know now that water and ice are competitive as a detector medium. They are, in fact, complementary. Water and ice seem to have similar attenuation length, with the role of scattering and absorption reversed. As demonstrated with the shallow AMANDA strings\cite{porrata}, scattering can be exploited to range out the light and perform calorimetry of showers produced by electron-neutrinos and showering muons. Long scattering lengths in water may result in superior angular resolution, especially for the smaller, first-generation detectors. This can be exploited to reconstruct events far outside the detector in order to increase its effective volume. 

\section*{Acknowledgements}

This research was supported in part by the U.S.~Department of Energy under Grant No.~DE-FG02-95ER40896 and in part by the University of Wisconsin Research Committee with funds granted by the Wisconsin Alumni Research Foundation.

\end{document}